\documentclass[aps,pre,twocolumn,superscriptaddress]{revtex4}

\usepackage{amsmath,amssymb,amsfonts}
\usepackage{graphicx}
\usepackage{dcolumn}
\usepackage{bm}
\usepackage{textcomp}
\usepackage{color}

\begin{document}

\title{Exact modeling method for discrete finite metamaterial lens}

\author{M. Lapine}
\affiliation{Dept.~Electronics and Electromagnetics, Faculty of Physics, University of Seville, 41015 Seville, Spain}

\author{L. Jelinek}
\affiliation{Department of Electromagnetic Field, Czech Technical University in Prague, 16627 Prague, Czech Republic}

\author{R. Marqu{\'e}s}
\affiliation{Dept.~Electronics and Electromagnetics, Faculty of Physics, University of Seville, 41015 Seville, Spain}

\author{M.\,J. Freire}
\affiliation{Dept.~Electronics and Electromagnetics, Faculty of Physics, University of Seville, 41015 Seville, Spain}

\email[Corresponding author:]{mlapine@uos.de}

\date{\today}

\begin{abstract}
We describe an efficient rigorous model suitable for calculating the properties of 
finite metamaterial samples, which takes into account the discrete structure of
metamaterials based on capacitively loaded ring resonators.
We illustrate how this model applies specifically to a metamaterial lens employed
in magnetic-resonant imaging.
We show that the discrete model reveals the effects which can be missed by a continuous
model based on effective parameters, and that the results are in close agreement with 
the experimental data.
\end{abstract}

\maketitle

%% =======================================================================
%% =======================================================================

\section{Introduction}

For the last decade, metamaterials \cite{SolSha,MarMarSor} are in the focus 
of research attention in theoretical and applied electrodynamics. 
Even though no commonly accepted definition is available \cite{ML7,Sih07},
this research direction experiences a boom encompassing a wide span of areas
ranging from microwave engineering to nonlinear optics.
One of the well-known suggestions for applications was formulated as a 
``perfect lens'' \cite{Pen00}, making use of negative effective material
parameters and providing imaging with subwavelength resolution.
The idea of super-resolution was subsequently analysed and developed in a number of ways
\cite{SKR1,MTA4,MFM5},
and even realized in practice (speaking about three-dimensional systems)
using split-ring resonators (SRRs) \cite{Fre5,FreMarJel08},
or transmission-line networks \cite{Grb5,AliTre07}.

Arguably the closest approach to practice offered by metamaterials, 
is related to magnetic resonance imaging (MRI).
For example, rotational resonance of magnetoinductive waves \cite{SolZhuSyd06}
was suggested for parametric amplification of MRI signals \cite{SymSolYou07}
or enhanced detection with flexible ring resonators \cite{SymYouSol10}.
Alternatively, applications based on `swiss-rolls' \cite{WHP3}
or wire media \cite{IkoBelSim06} channelling
were put forward \cite{WHP3,RadGarCra09}.
Naturally, superlens concept is also promising in this area:
specifically for MRI, an isotropic lens 
based on capacitively loaded single ring resonators
was designed and experimentally tested \cite{FreMarJel08}.

Such a metamaterial lens is intended
to operate at the value of effective permeability $\mu=-1$.
In theory, for modelling such metamaterials (based on SRRs), one can exploit an
effective medium approach, taking care of numerous limitations related
to general restrictions of homogenization \cite{Agr09}
as well as to specific peculiarities caused by resonant nature
of the structural elements \cite{Sim08}.
Universally, all the structure details (size of the elements and 
lattice constants) must be much smaller than the wavelength; while
the total number of elements in metamaterial should be sufficiently
large to make homogenization meaningful.
In addition, spatial dispersion effects can be rather remarkable
in metamaterials, and impose further restrictions on effective
medium treatment, prohibiting that in certain frequency bands \cite{Sim08}.

Unless one opts for a completely numerical homogenization 
method \cite{Sil07}, generally applicable to almost arbitrary structures,
a model have to be developed to describe adequately the effective medium 
properties.
Quite general approach \cite{Sim08} for homogenization of resonant metamaterials
can be applied to a variety of metamaterials including those which combine
different element types. 
However, this relies on the dipole approximation for the interaction of
elements in the lattice, which may not be always valid.
For example, mutual interaction of the circular currents close to each other
significantly differs from dipole interaction, which becomes relevant for
dense metamaterials.
The first rigorous analysis of such metamaterials was given early in Ref.~\cite{GLS2},
where the effective permeability has been derived given the properties of individual
elements and lattice parameters, through the classical procedure of averaging
the microscopic fields. 
In that approach, mutual inductances between a large number of neighbours 
are taken into account, revealing the importance of lattice effects.
This approach, however, is limited to quasi-static conditions, and requires
wavelength to be much larger than any structural details.
Later on, a rigorous method was elaborated for isotropic lattices of resonant
rings \cite{BaeJelMar08}, which accounts for spatial dispersion. 
On the other hand, the latter approach employs a nearest neighbour approximation
as otherwise full analytical treatment becomes impractical.

The above theoretical methods provide the effective parameters for ``infinite'' structures 
(which in practice implies the structures sufficiently large in all three dimensions).
The lens of Ref.~\cite{FreMarJel08}, however, contains just a few elements across the slab.
Specifically for this case, a method was developed to calculate the transmission properties 
for a thin infinite slab \cite{JelMarFre09}; furthermore, it was shown that similar results 
can be obtain by considering an equivalent slab with the 
effective medium parameters as obtained in Ref.~\cite{BaeJelMar08}.

Nevertheless, it is clear that a number of peculiar effects caused by the 
discrete structure of the lens as well as its finite size, cannot be 
reliably assessed with the above models, as the lens is too small for 
an effective medium treatment.
On the other hand, it is large enough to make an analysis with full-wave
commercial software practically impossible.
For this reason, here we develop a finite model to calculate lens properties, 
which explicitly takes all the structural details into account.
The goal of this paper is to describe this modelling approach in detail,
and to illustrate that indeed it does reveal some features which are
missed by the continuous modelling.
We should note, though, that while the model is described in connection 
to one particular structure, the approach applied here is generally 
applicable to any realistic SRR-based metamaterial, and therefore 
is useful for a wide range of applications.

%%%%%%%%%%%%%%%%%%%%%%%%%%%%%%%%%fig.1%%%%%%%%%%%%%%%%%%%%%%%%%%%%%%%%%%%%%%%%%
\begin{figure}[b]
\centering
\includegraphics{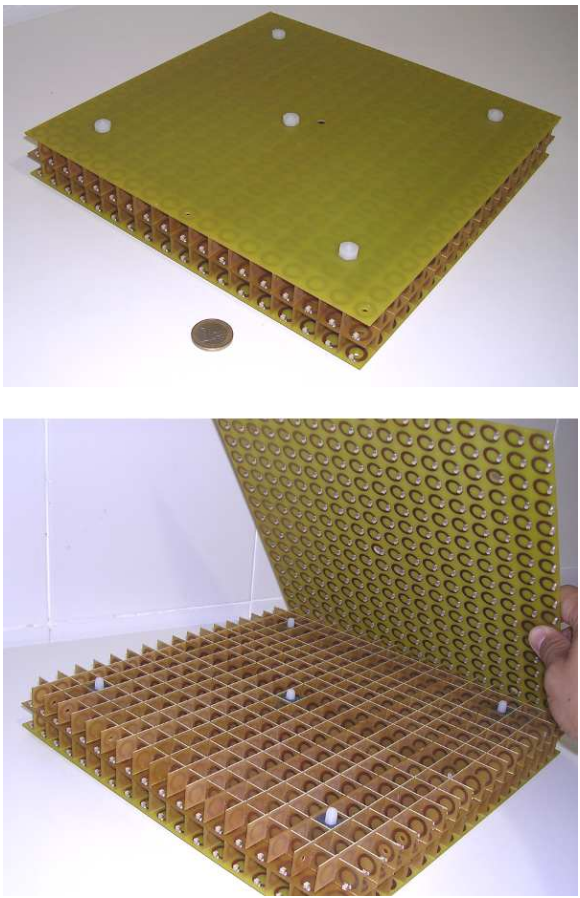}	%%% [width=1\columnwidth]
\caption{\label{lensfoto}Photograph of the quasi-magnetostatic metamaterial lens analysed in this paper.}
\end{figure}

\section{Geometry of the problem}

The metamaterial lens described in Ref.~\cite{FreMarJel08} is composed of 
capacitively loaded rings (CLRs) periodically arranged in an isotropic 
three-dimensional lattice with the lattice constant $a = 1.5$\,cm. 
The lens features three planes of 18 by 18 CLRs interlayered with orthogonal
segments providing two mutually orthogonal sets of two layers 17 by 18 CLRs each
(see Fig.~\ref{lensfoto} for clarity), which makes it up to roughly 2200 CLRs.
This lens can be optionally extended by an extra 3D-layer, resulting 
in having four $18 \times 18$ layers interlaced with the two orthogonal
subsystems of 3 by 17 by 18 CLRs, amounting to about 3130 elements.
Overall dimensions of the (non-extended) lens are thus 
$27 \times 27 \times 3$~cm.

\begin{figure}[t]
\centering
\includegraphics[width=0.9\columnwidth]{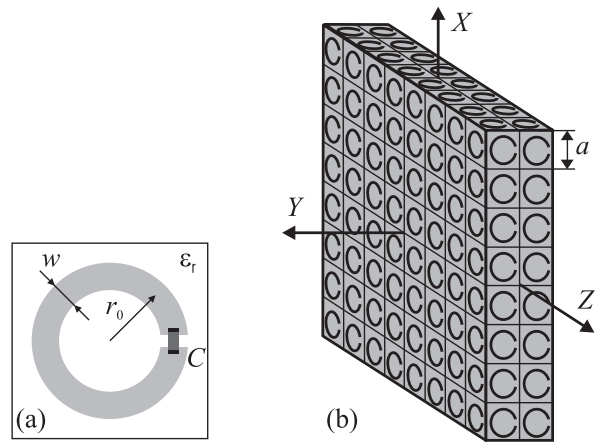}
\caption{\label{lensgeom} (a) Sketch of the CLR resonator; 
(b) Scheme of the lens with the corresponding coordinate system.}
\end{figure}

The CLRs themselves (Fig.~\ref{lensgeom}a) are made of copper 
through etching metallic strips on a dielectric board. 
The mean radius $r_0$ of the CLRs is 0.49\,cm ($r_0 / a = 0.33$) and 
the strip width $w$ is 0.22\,cm ($w / a = 0.15$). 
The CLRs are loaded with lumped non-magnetic 470\,pF capacitors. 
The self-inductance of the CLRs, $L =  \omega^2_0 / C = 13.5$\,nH, 
has been obtained from the measured value of the frequency of resonance in free space, 
equal to 63.28\,MHz ($k_0 a = 0.02$).  
By measurement of the quality factor of the resonator the resistance 
has been estimated as $R = 0.0465$\,Ohm, 
which includes the effects of both the ring and the capacitor.

%%%%%%%%%%%%%%%%%%%%%%%%%%%%%%%%%fig.2%%%%%%%%%%%%%%%%%%%%%%%%%%%%%%%%%%%%%%%%%

We reserve the standard coordinate system ($x$, $y$, $z$) for discussions
on the level of geometry of one ring and their mutual interactions, 
as relevant for the next section. 
When referring the overall lens geometry, we define supplementary coordinate 
system ($X$, $Y$, $Z$) so that the lens geometrical 
centre is placed at the coordinate origin, and the $Y$ axis is perpendicular 
to the lens as slab (``lens axis''), while the long edges of the lens are 
parallel to $X$ and $Z$ axes (Fig.~\ref{lensgeom}b). 
Note that the coordinate origin is between the rings, and thus the lens 
is completely symmetric with respect to the coordinate origin, all the axes 
and all the coordinate planes
(in the analysis, we neglect minor asymmetry occurring in the real lens 
caused by specific assembly details, e.g. resulting from substrate thickness,
as these deviations are of the same order as unavoidable production inaccuracy).
For the three mutually orthogonal sets of CLRs, we will refer to as
X-rings, Y-rings or Z-rings, depending on whether the rings'
normals are along $X$, $Y$ or $Z$ axes.
The lens, therefore, contains 612 of either X-rings and Z-rings,
and 972 Y-rings.
We will also introduce a consecutive numbering of all the CLRs with a single index.
The so called input and output surfaces of the lens correspond to $Y = \mp 1.5$\,cm,
while the theoretical source and image planes are at $Y = \mp 3.0$\,cm.

\section{Theoretical model}

For the analysis of the lens response to the external field, 
we consider an ideal cubic lattice of L--C circuits supporting current. 
With the time convention as
$I \propto\exp({\mathrm j} \omega t)$, 
each of the currents is governed by equation
\begin{equation}
  \label{eq1}
  Z_0 I_n = -{\mathrm j} \omega\Phi_n,
\end{equation}
where the self-impedance 
$Z_0 = \left( R + {\mathrm j} \omega L + 1 / \left( {\mathrm j} \omega C \right) \right)$ 
is determined by the resistance $R$, self-inductance $L$ and self-capacitance $C$
of the single CLR, while $\Phi_n$ represents the total magnetic flux through the 
considered ring which can be written as
\begin{equation}
  \label{eq2}
\Phi_n  = \Phi_n^{\text{ext}} + \sum\limits_{ m \ne n } {\Phi_{nm}} 
= \Phi_n^{\mathrm {ext}} + \sum\limits_{ m \ne n } {M _{nm} I_m }
\end{equation}
where $\Phi_n^{\rm {ext}}$ is the magnetic flux from external sources 
and $M _{nm}$ are the mutual inductances between the rings $n$ and $m$. 
Combining Eq.~\eqref{eq1} and Eq.~\eqref{eq2} we obtain
\begin{equation}
\label{impsyst}
\mathbf{\overline Z} \cdot \mathbf{I} =  - {\mathrm j} \omega {\mathbf{\Phi}}^{\mathrm{ext}} 
\end{equation}
with $Z_{nn} = Z_0$, $Z_{nm} = \mathrm j \omega M_{nm}$,
which is a system of linear equations for unknown currents, provided that 
the external sources are known.

Mutual inductance between the flat rings (which is the case under consideration)
carrying the currents $I_n$ and $I_m$ uniform along the ring contour
is, most generally \cite{Landau}, 
\begin{equation}
  \label{surfmutl}
M_{nm} = \frac{\mu}{4 \pi I_n I_m} 
\int\limits_{S}\!\int\limits_{S'} 
\frac{\mathbf{K}_n (\mathbf{r}) \cdot \mathbf{K}_m (\mathbf{r'})}%
{\left| \mathbf{r} - \mathbf{r'} \right|} \: \mathrm{d}S \, \mathrm{d}S' 
\end{equation}
where $\mathbf{K}$ represent surface current densities; we assume that these 
follow Maxwellian distribution across the strip,
\begin{equation}
  \label{maxw}
K_\varphi (\rho) = \frac{2I}{w\pi 
\sqrt{ 1 - \left( \dfrac{\rho  - r_0}{w/2} \right)^2 } },
 \quad
 \int\limits_{r_0  - w/2}^{r_0  + w/2} {K_\varphi \: \mathrm{d}\rho } = I. 
\end{equation}

Clearly, such integration is not ideally suited for numerical calculation.
In the first approximation, mutual inductance between CLRs can be
estimated with the one between linear currents 
(double linear integration along the equivalent ring contour),
but for close CLRs this does not give a good precision. 

However, a trick is that the result of surface integration according to Eq.\,\eqref{surfmutl}
can be approximated with a good precision through an average mutual inductance 
between two pairs of circular currents \cite{RosaGrover}. 
This way, each flat ring can be represented by a pair of coaxial circular currents 
of radii $r_0 \pm \gamma w / 2$, 
and the sought mutual inductance is calculated as an average between 
the four corresponding linear ones:
\begin{equation}
M_{nm} = \left( L^{++}_{nm} + L^{--}_{nm} + L^{+-}_{nm} + L^{-+}_{nm} \right) / 4, \label{avgmutl}
\end{equation}
which essentially decreases calculation time.
The value of particular parameter $\gamma$ depends on the ring geometry, but does not depend
remarkably on the relative orientation and distance between the CLRs (within the limits of
lens structure). For the particular parameters considered here, 
$\gamma \approx 0.7$ was numerically found to give a good match to the precise
integration \eqref{surfmutl} (while $\gamma = 1$ would correspond to the edges of the strip).

%%%%%%%%%%%%%%%%%%%%%%%%%%%%%%%%%fig.3%%%%%%%%%%%%%%%%%%%%%%%%%%%%%%%%%%%%%%%%%
\begin{figure}[t]
\centering
\includegraphics[width=1\columnwidth]{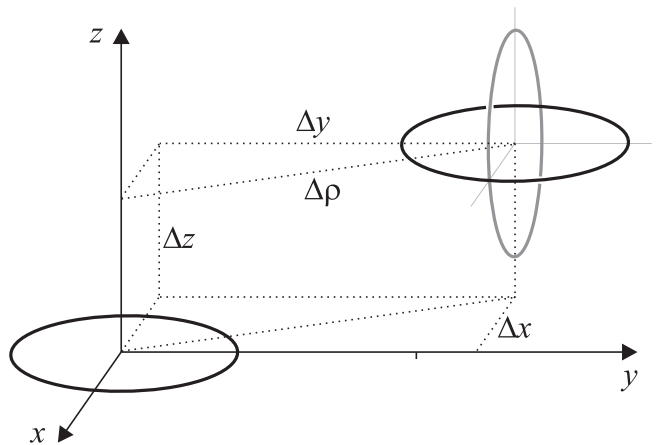}
\caption{\label{mutuals} Geometry of the linear currents for parallel or orthogonal ring orientations, 
as relevant for mutual inductance calculations.}
\end{figure}

To achieve a faster calculation for the ``linear'' mutual inductance itself, 
we further note that it can be easily
evaluated \cite{Landau} by integrating
the vector potential $\mathbf{A}$ along the current contour $l$:
\begin{equation}
L_{nm} = \frac{1}{I_n} \int\limits_{l_m} \mathbf{A_{\mathit{n}}} \cdot \mathbf{d l_{\mathit{m}}}, \label{linmutl}
\end{equation}
with vector potential itself having only $A_{\varphi}$ component
(assuming that the source ring is placed at the coordinate origin 
with the normal along $z$ axis, see Fig.~\ref{mutuals}),
which can be obtained with the help of elliptic integrals as \cite{Landau}
\begin{equation}
A_{\varphi} = I \frac{\mu_0}{4 \pi} \sqrt{\frac{r_0}{\rho}}
\left( \frac{ \left( 2 - \kappa^2 \right) \mathcal K(\varkappa) -  2 \mathcal E(\varkappa) }{\varkappa} \right), 
\end{equation}
with
\begin{equation}
\varkappa = \sqrt{\frac{4 r_0 \rho}{(\rho + r_0)^2 + z^2}} \notag
\end{equation}
being the argument of complete elliptic integrals 
\begin{equation*}
\mathcal E = \int\limits_0^{\pi/2} \sqrt{1 - \varkappa^2 \sin^2 \theta} \; \mathrm{d}\theta, \qquad 
\mathcal K = \int\limits_0^{\pi/2} \frac{\mathrm{d}\theta}{\sqrt{1 - \varkappa^2 \sin^2 \theta}}. 
\end{equation*}
Given the fact that the fast pre-defined routines for elliptic integrals are available in
a number of computational platforms (e.g.\ Matlab$^{\circledR}$), this effectively
reduces double integration to a single one.
Thus, finally, only four linear integrals like \eqref{linmutl} are required to approximate the exact 
value of a double integration like \eqref{surfmutl}.

\section{Numerical implementation}

To analyse the response of the lens to an external field source, the key step
lies in solving system \eqref{impsyst}. To do so, we need to know the matrix 
of mutual inductances $\mathbf{\overline M}$. This matrix is only determined
by the geometry of the rings arrangement inside the lens and can be calculated
once for a given lens geometry, while the impedance matrix $\mathbf{\overline Z}$ 
can be then obtained for all frequencies as shown after \eqref{impsyst}.

For such a lens as described above, having 2196 rings, the matrix contains
almost 5 million values, and filling those with a direct calculation would be rather
time-consuming even with a simplified integration described in the previous section.
However, obvious reciprocity ($M_{nm} = M_{mn}$) and symmetry properties of the lens 
allow for a great simplification of matrix filling. 
Indeed, the lens is symmetric with respect to $X$, $Y$ and $Z$ axes 
as well as to $XY$, $YZ$ and $ZX$ planes. 
This implies, in particular, that the mutual inductances between X-rings and Y-rings
are all the same as between Z-rings and Y-rings.
Furthermore, as all the rings are identical, inductance between them 
is only determined by their mutual orientation and spatial offsets 
$\Delta x$, $\Delta y$, $\Delta z$ (see Fig.~\ref{mutuals}), 
and for parallel rings even $\Delta x$ is equivalent to $\Delta y$.
Explicitly, integration for the mutual inductances between the parallel
rings is performed according to 
\begin{gather}
L^{\text{P}} (\Delta b, \Delta z) = \int\limits_{0}^{2\pi}
A_{\varphi} \frac{r_2 (r_2 + \Delta b \cos \alpha)}{\Delta \rho} \; \mathrm d \alpha, 
\\ 
\Delta \rho = \sqrt{r_2^2 + (\Delta b)^2 + 2 r_2 \Delta b \cos \alpha}, \notag
\\
\varkappa^2 = {\frac{4 r_1 \Delta\rho}{(\Delta\rho + r_1)^2 + (\Delta z)^2}} \notag
\label{pmutl}
\end{gather}
where $\Delta b = \sqrt{(\Delta x)^2 + (\Delta y)^2}$; and for the rings with 
orthogonal mutual orientation
\begin{gather}
L^{\text{O}} (\Delta x, \Delta y, \Delta z) = \int\limits_{0}^{2\pi}
A_{\varphi} \frac{r_2 \Delta y \cos \alpha}{\Delta \rho} \; \mathrm d \alpha,
\\
\Delta \rho = \sqrt{(\Delta x - r_2 \sin \alpha)^2 + (\Delta y)^2}, \notag
\\
\varkappa^2 = {\frac{4 r_1 \Delta\rho}{(\Delta\rho + r_1)^2 + (\Delta z - r_2 \cos \alpha)^2}}. \notag
\end{gather}
In the above equations, we imply a general case that the radii
of the two rings ($r_1$ and $r_2$) can be different.

Thus, a number of ring pairs within the lens share the same value
of mutual inductance, so it is only necessary to calculate a full set of
non-equivalent mutuals and then assign those values depending on the mutual
offsets.
With the particular lens considered here, there are only 1924 independent inductances
for the parallel ring orientation, and 1668 for the orthogonal one, so the total
number of calculations \eqref{avgmutl} is 3592 --- orders of magnitude smaller than 
the number of matrix elements.
This way, the entire matrix can be filled in a matter of seconds on an ordinary PC.

Another preliminary step is to determine the external flux $\mathbf \Phi^{\text{ext}}$ 
imposed to each ring by a given  source. 
For a homogeneous field or a plane wave excitation, calculation 
is straightforward with the known coordinates of each ring: 
\begin{equation}
\Phi_n^{\text{ext}} = \pi r_0^2 \, \mathbf B_n \cdot \mathbf n,
\end{equation}
where $\mathbf n$ is a ring normal while magnetic field $\mathbf B_n$ can be evaluated
at the ring centre as the field variation across the ring is negligible.

%%%%%%%%%%%%%%%%%%%%%%%%%%%%%%%%%fig.4%%%%%%%%%%%%%%%%%%%%%%%%%%%%%%%%%%%%%%%%%
\begin{figure}[b]
\centering
\includegraphics[width=0.95\columnwidth]{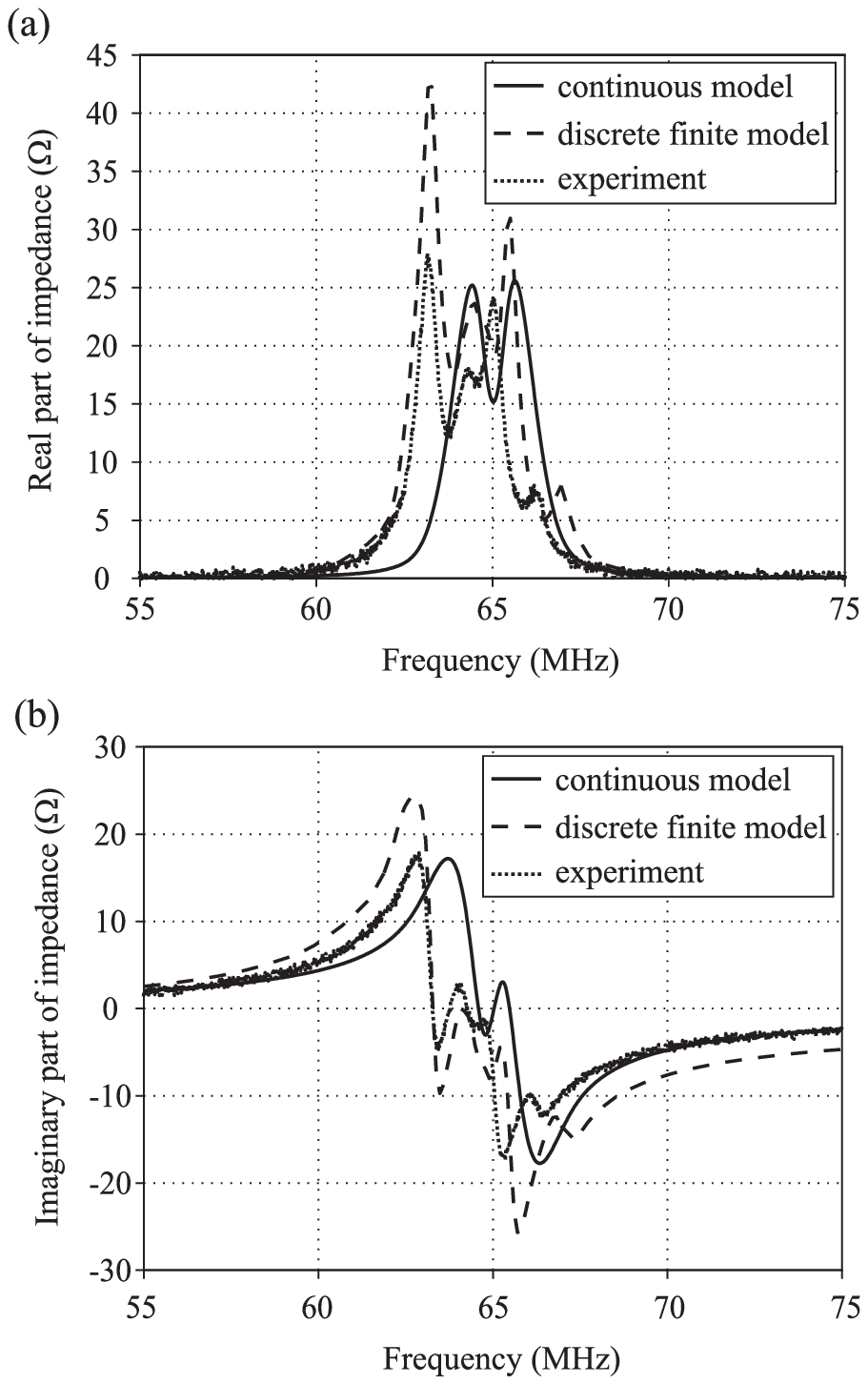}
\caption{\label{imput} Frequency dependence  of the real (a) and imaginary (b) parts 
of the impedance measured by a 3-inch coil placed at the image plane ($Y = -3$\,cm).} 
\end{figure}

In practice, the lens is typically used along with excitation / measuring coils
employed in MRI practice. 
In that case, instead of calculating the field produced by a coil over each ring 
(which is further complicated as this field is not uniform across the ring), 
it is much easier to obtain the flux directly
\begin{equation}
\Phi_n^{\text{ext}} = M^{\text{c}}_{n} I^{\text{c}}
\end{equation}
in terms of mutual inductance $M^{\text{c}}_n$ between the coil and each ring, 
which can be calculated with the same method as the one between the rings.
Above, $I^{\text{c}} \equiv I_\text{N+1}$ is the total current induced in the coil by the
external voltage source as well as by the lens.
Imposing a given voltage $V_{\text{c}}$ to the coil with the self-impedance $Z_\text{c}$, 
we can include the coil mutual impedances into system \eqref{impsyst}, modified as
\begin{gather}
\label{coilsyst}
\mathbf{\overline Z} \cdot \mathbf{I} =  \mathbf{V}  
\\
Z_{nn} =
\begin{cases}
Z_0 \\
Z_\text{c} 
\end{cases}
Z_{nm} =
\begin{cases}
\mathrm j \omega M_{nm} & \text{for} \quad 1 \leqslant n \leqslant N\\
\mathrm j \omega M^{\text{c}}_n & \text{for} \quad n = N+1
\end{cases} \notag
\end{gather}
with $V_n = V_{\text{c}} \, \delta_{n,N}$ and $N$ being the total number of rings in the lens.
Clearly, additional coils, if necessary, can be included by extending the matrix system 
in an analogous way.

After the above procedures, it is finally possible to solve the systems 
\eqref{impsyst} or \eqref{coilsyst} obtaining currents  $I_n$ in each ring for any given excitation.
With these known, it is further possible to calculate any desired response of the lens,
such as magnetic field produced by the lens (using standard Biot-Savart expressions) 
or impedance as measured by the MRI coil,
\begin{equation}
Z^\text{coil} = \sum\limits_{n=1}^{N} \mathrm j \omega M^{\text{c}}_{n} \: \frac{I_n}{I^{\text{c}}}.
\label{impbycoil}
\end{equation}

\section{Results and discussion}

Armed with the above precise method, we can have a detailed look into lens features 
and response to various external field sources.
In previous work \cite{JelMarFre09} it was concluded that the accurate model, 
developed for an 2D-infinite slab with the same structure and thickness as
the real lens, is capable of predicting the observations made in connection
to lens use in MRI practice.
In a typical setup, a coil of 3 inch in diameter is placed
parallel to the lens interface at the source plane, $Y=-3$\,cm 
(that is, at a distance 1.5\,cm, equal to one half of the lens thickness from the lens surface).
The super-lens behaviour implies that the magnetic field produced by the coil, is then 
reproduced in the space behind the image plane ($Y=3$\,cm), as if the coil itself were 
present in place of the image.

%%%%%%%%%%%%%%%%%%%%%%%%%%%%%%%%%fig.4%%%%%%%%%%%%%%%%%%%%%%%%%%%%%%%%%%%%%%%%%
\begin{figure}[t]
\centering
\includegraphics[width=0.95\columnwidth]{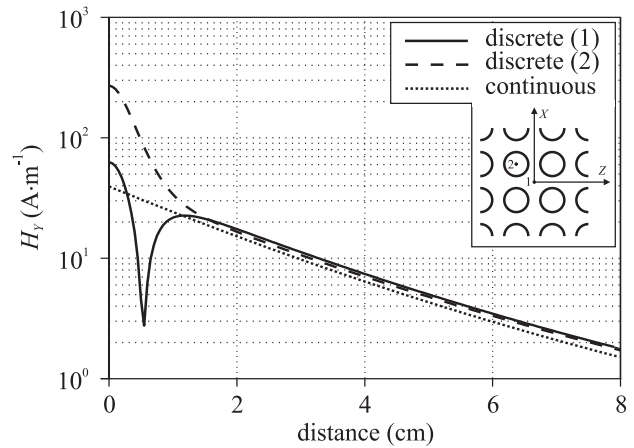}
\caption{\label{hfcomp} Axial component $H_Y$ of the total magnetic field observed behind the lens
surface along the lens axis (1) or along the parallel line (2) slightly displaced in $X$ and $Z$
direction so that it passes through the centre of one ring (see the inset for the labels of the axes).
Comparison between the two models when the lens is excited by 3-inch coil, 
centred with respect to the lens axis and placed at $Y = -1.5$\,cm.}
\end{figure}

A straightforward example to test the developed model and to compare it with practice
as well as with earlier models, is to evaluate the impedance as measured by a coil 
in front of the lens, depending on frequency. 
In the discrete model, it is given by Eq.~\eqref{impbycoil},
while with the continuous model 
(for an infinite homogeneous slab with an appropriate effective permeability \cite{BaeJelMar08})
it can be numerically calculated as 
\begin{equation}
Z^\text{coil}  = - \dfrac{1}{I^{\text c}} \: \mathsf{Re} \int\limits_\text{coil}
\mathbf{E}^{\text r} \cdot \mathbf{d l}^{\text c},  
\end{equation}
where $\mathbf{E}^{\text r}$ is electric field reflected by the lens.
The two modelling results are compared in Fig.~\ref{imput} along with the experimental data.
Although there is no exact quantitative matching to the measured data, it is clear
that the frequency dependence provided by the discrete model is closer to experiment than that of the continuous calculation.
On the other hand, we can conclude that the latter already provides qualitatively
suitable picture, predicting an overall pattern of the impedance frequency dependence.

With both the continuous model \cite{JelMarFre09} and the model developed above,
it is easy to obtain the axial magnetic field $H_Y$ behind
the lens for a given excitation.
Comparison between the predictions of the two models is shown in Fig.~\ref{hfcomp}.
One can see that at distances smaller than about one lattice constant ($a=1.5$\,cm),
$H_Y$ is essentially inhomogeneous as the near-field of the individual rings dominates,
so that the total field is quite different whether traced along the lens axis 
(which passes between the rings) or along a line that passes through a ring centre,
while both are remarkably different from the continuous model. 
This is an obvious consequence of the discrete lens structure, which cannot
be revealed by a homogenized model but is apparent in practice.
At distances larger than approximately one lattice constant ($1.5$\,cm), 
the field observed along the two axes converge, and are qualitatively
similar to the continuous model with a fair numerical agreement 
(see Fig.~\ref{hfcomp}).

%%%%%%%%%%%%%%%%%%%%%%%%%%%%%%%%%fig.4%%%%%%%%%%%%%%%%%%%%%%%%%%%%%%%%%%%%%%%%%
\begin{figure}
\centering
\includegraphics[width=1\columnwidth]{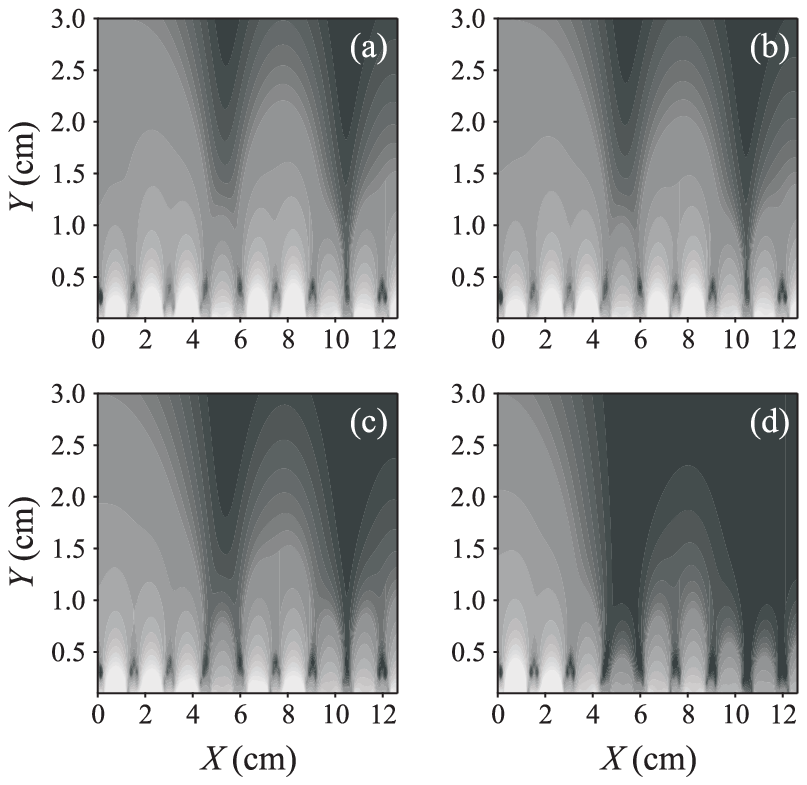}
\caption{\label{varcoil} Axial component $H_Y$ of the magnetic field observed behind the lens
surface. Horizontal axis corresponds to the lens surface (parallel to $X$--$Z$ plane),
while the vertical one is normal to the lens ($Y$).
Only one half of the symmetrical field spatial distribution is presented;
normalized magnitudes are shown in logarithmic scale between 10 (white) and 0.1 (black) A/m.   
Excitation with coils of different radii (0.5\,cm, 1.5\,cm, 3\,cm and 4.5\,cm), 
centred with respect to the lens axis and positioned at $Y = -1.5$\,cm.}
\end{figure}

Another peculiarity arising from the discrete structure is related to the spatial 
resolution of the lens in the $X$--$Z$ plane. 
Evidently, a lens cannot resolve any details which are separated by distances
of the order of lattice constant. 
To identify the actual limitation, we test the magnetic field distributions 
originating from using the coils of various small radii (Fig.~\ref{varcoil}).
For excitations with a coil of the ring size, the entire lens is dominated by 
standing magnetoinductive waves \cite{SKR2b}, and the field pattern does
not suggest any hints for resolving the source (Fig.~\ref{varcoil}a).
Indeed, practically the same field pattern is observed with a three times
larger coil, two lattice constants in diameter (Fig.~\ref{varcoil}b).
With a still larger coil, encompassing four lattice constants, one may 
argue that the pattern starts to clarify (Fig.~\ref{varcoil}c), 
though still it cannot be reliably used to assess the source location and size.
A reasonable picture is obtained for a 4.5\,cm coil radius, where the field
farther than the image plane looks as expected with super-lens performance (Fig.~\ref{varcoil}d).
We can therefore conclude that spatial resolution of the discrete lens 
can be assumed to be of the order of 5 lattice constants. 
This observation is in good agreement with the general concerns regarding
the lattice effects in metamaterials \cite{GLS2}.

With the above examples, we clearly demonstrate that the exact model described 
in this paper, is suitable for a reliable description of the metamaterial lens,
and makes it possible to predict specific observations which might be missed by
a continuous model.

Certainly, the above methodology is not restricted to the 
particular lens geometry and can be perfectly used for any metamaterials 
designed with CLRs or SRRs, whether isotropic or anisotropic, and also 
arbitrarily small in size. 
The only limitation is that for very large number of elements, numerical 
evaluation on conventional computers may fail, specifically as far as
allocating space for huge impedance matrices, and inverting these, is concerned.
However, in metamaterial research it rarely comes to samples that large, and, 
on the other hand, when it comes, then there are good reasons to expect that
continuous models will work sufficiently fine. 

In contrary, for small metamaterials typically considered for practical use, 
modelling this way provides an invaluable insight into their properties and
leads to reliable predictions.

\begin{acknowledgements}
This work has been supported by the Spanish Ministerio de Educaci\'on y Ciencia and
European Union FEDER funds (projects TEC2007-65376, TEC2007-68013-C02-01, and
CSD2008-00066), by Junta de Andaluc{\'\i}a (project TIC-253), and by Czech Grant Agency
(project no.\ 102/09/0314).
\end{acknowledgements}

%% =======================================================================
%% =======================================================================

% \newpage

%% =======================================================================

\end{document}